\documentclass{article}
\usepackage{graphicx}
\usepackage{authblk}
\usepackage{verbatim}
\usepackage{multirow}
\usepackage{array, tabularx}
\usepackage{subfigure}
\usepackage{tikz-network}

\usepackage{hyperref} 

\usepackage{comment}

\usepackage[a4paper, total={6.6in, 8in}]{geometry}
\usepackage{xr}
\usepackage[utf8]{inputenc}
\usepackage[T1]{fontenc}
\usepackage{amsthm}
\usepackage{amsmath}
\usepackage{amssymb}
\usepackage{siunitx}

\usepackage{url}
\usepackage{booktabs} 
\usepackage{xspace}
\usepackage{float}
\usepackage{multirow}
\usepackage{alltt}
\usepackage{color}
\usepackage{pifont}
\usepackage{makecell}
\usepackage{multicol}
\usepackage[super]{nth}

\let\origref\ref
\def\ref#1{\textnormal{\origref{#1}}}

\title{Slaying the Dragon: The Quest for Democracy in Decentralized Autonomous Organizations (DAOs)}

\author[1,*]{Stefano Balietti, 0000-0002-6056-6765}
\author[2,3,*]{Pietro Saggese, 0000-0002-7387-0846}
\author[3]{Stefan Kitzler, 0000-0001-8454-167X}
\author[3]{Bernhard Haslhofer, 0000-0002-0415-4491}

\affil[1]{University of Mannheim, Mannheim, DE. E-mail: stefano.balietti@uni-mannheim.de}

\affil[2]{IMT Scuola Alti Studi Lucca, Lucca, IT. E-mail: pietro.saggese@imtlucca.it}

\affil[3]{Complexity Science Hub, Vienna, AT. E-mails: {kitzler,haslhofer}@csh.ac.at}

\date{} 

\begin{document}
	\maketitle
	
	\begin{abstract}
        This chapter explores how Decentralized Autonomous Organizations (DAOs), a novel institutional form based on blockchain technology, challenge traditional centralized governance structures. DAOs govern projects ranging from finance to science and digital communities. They aim to redistribute decision-making power through programmable, transparent, and participatory mechanisms. This chapter outlines both the opportunities DAOs present, such as incentive alignment, rapid coordination, and censorship resistance, and the challenges they face, including token concentration, low participation, and the risk of de facto centralization. It further discusses the emerging intersection of DAOs and artificial intelligence, highlighting the potential for increased automation alongside the dangers of diminished human oversight and algorithmic opacity. Ultimately, we discuss under what circumstances DAOs can fulfill their democratic promise or risk replicating the very power asymmetries they seek to overcome.
		\let\thefootnote\relax\footnotetext{$^*$ Corresponding authors: Stefano Balietti, Chair of Data Science in the Economic and Social Sciences, University of Mannheim, Mannheim, Germany. Pietro Saggese, Assistant Professor, Scuola IMT Alti Studi Lucca, Italy.}

\end{abstract}

\section{Introduction}

Human institutions have a natural tendency towards centralization. Across time and domains, the concentration of decision-making power is a recurring pattern. 
Historically, the tendency toward centralization is well-documented: ancient empires such as Rome and China centralized power to enforce order and stability over vast territories, often at the expense of local autonomy and participatory governance \cite{hoyt1941forces,hopkins1965elite,ko2013regional}. In the modern era, centralization has become a multi-headed dragon. Politically, it manifested itself in the 19th century with the rise of nation-states with strong executive branches \cite{wimmer2010rise}. Economically, multinational corporations with complex hierarchical structures have progressively dominated the markets over the last hundred years \cite{kwon2024100}. Even in science, centralization has become an issue: a few powerful publishing houses control the vast majority of outlets \cite{trueblood2025misalignment}; scientists from elite universities have disproportionate access to research funds and, hence, more power to shape entire scientific fields \cite{merton_1968_matthew,xie_undemocracy_2014}. More broadly, the tendency towards centralization is a realization of scaling laws that have been empirically measured in all kind of human systems: from cities \cite{bettencourt2007growth} to firms \cite{gabaix2009power}, from accumlated family wealth \cite{sinha2006evidence} to links in the World Wide Web \cite{adamic2000power}.

If centralization is ubiquitous, it is perhaps because it has an evolutionary advantage in key dimensions such as coordination, efficiency, and scalability \cite{diamond1999guns}. For instance, the Roman Empire’s centralized bureaucracy allowed it to collect taxes, maintain roads, and enforce laws across thousands of miles. In the corporate world, centralized giants like General Motors in the 20th century and Amazon today have leveraged hierarchical command structures to rapidly scale operations, optimize logistics, and control global supply chains. In science, large-scale, coordinated research efforts---such as the Human Genome Project or CERN's particle physics experiments---were possible thanks to centralized planning and pooled investment. 
Along these lines, centralization is currently emerging within artificial intelligence (AI) systems. The training of foundational models is now concentrated in the hands of a few key actors with sufficient technical expertise, and whose dominance is reinforced by the need for massive capital requirements - in the order of billions of dollars - to build AI infrastructure. 

In general, centralization enables the allocation of resources with purpose and precision, particularly in times of crisis, when quick, top-down decisions often outperform fragmented efforts \cite{rosenthal2001managing}. However, the benefits of centralization quickly vanish when institutions---in the words of Nobel laureate Daron Acemoglu---switch from being ``inclusive'', where power is distributed broadly and institutions foster innovation and prosperity, to being ``extractive'', with power and wealth concentrated within a small elite \cite{robinson2012nations}. 
The evolution of AI reflects comparable tensions between concentration of power and the potential for broader inclusion. 
Agentic AI approaches, i.e., systems that can automate complex workflows including decision-making processes, are growing rapidly; we can reasonably anticipate that in the near future, AI systems will be able to make decisions with minimal human 
oversight while pursuing predefined objectives.
However, AI decision-making remains largely opaque and cannot be attributed to any accountable individual or entity, raising concerns about accountability, power asymmetries, and the legitimacy of outcomes. 
Therefore, depending on how they are designed, future AI-based decision-making systems can either reinforce hierarchical control and concentration of power, or enhance coordination and support inclusive, participatory governance. 
Understanding the problems with centralization is therefore crucial for designing governance systems that remain inclusive and resilient.

\subsection{The problems with centralization}

In top-down centralized institutions power usually concentrates in the hands of a few and internal decision-making processes are inaccessible to stakeholders. This combination of features can be origin to a number of important drawbacks which can be conceptualized in five main categories (see Table \ref{tab:problems_of_centralization}). First of all, opaque decision processes weaken (i) \textit{accountability}, favoring risky behavior and moral hazard; even worse, they create fertile ground for corruption. Second, even in absence of malevolent actors, centralized institutions may reach (ii) \textit{worse decision outcomes} because of the reduced participation of stakeholders to the decision-making; this in fact limits access to information and favors groupthink  \cite{hong_diversity_vs_experts_2004,kavadias_diversity_brainstorming_2009}. Third, centralized authorities may actively engage in discriminatory actions aiming to reduce (iii) \textit{inclusion}, for instance by censoring or persecuting real and potential challengers and minorities. Fourth, centralized institutions tend to ossify in obsolete governance structures because (iv) \textit{innovation} is stifled by the lack of incentives to pursue it \cite{moch1977size}; in this context, centralization inherently entails a trade-off with personalization and diversity \cite{brown1993centralized}. Finally, centralized systems in the form of hub-spoke networks are known to have limited (v) \textit{processing capacity} and to be prone to congestion \cite{bayram2023hub}, hindering the ability to fast respond to changes in the environment.

Altogether the above points may explain why centralized institutions are ``generally characterized by stability and incrementalism, but occasionally they produce large-scale departures'' \cite{true2019punctuated}. However, decentralized systems can also undergo abrupt changes and the parabola of the Internet is a spectacular testimony of how dramatic regime shifts toward centralization can be \cite{lanier2014owns}.

\begin{table}
    \centering
    \begin{tabular}{lp{0.2\textwidth}p{0.33\textwidth}p{0.33\textwidth}}
        & Category & Mechanism & Negative consequences \\
        \hline
        (i) & Accountability & Opaque decision processes & Moral hazard, corruption \\
         \hline
        (ii) & Outcome & Reduced participation, groupthink, difficulty gathering information & Worse decision outcomes \\
         \hline
        (iii) & Inclusion & Struggle to remain in power & Persecution of challengers and minorities, censorship, systemic distrust \\
         \hline
        (iv) & Innovation & Lack of incentives & Lack of innovation, diversity, customization \\
         \hline
        (v) & Capacity & Hubs as processing bottlenecks  & Congestion, slower response to quick changes \\
    \end{tabular}
    \caption{\textbf{The problems with centralization.} Authors' conceptualization in five categories.}
    \label{tab:problems_of_centralization}
\end{table}

In the words of Tim Berners-Lee, the inventor of the World Wide Web, the early Internet was ``a blank sheet of paper'' where ``no permission was needed to publish anything'' \cite{berners1999weaving}. In this fully decentralized environment,  grassroots communities were thriving: The WELL fostered free speech and collective intelligence; FidoNet linked global users through dial-up BBSes with a federated messaging system; IRC powered real-time, uncensored chat for developers and activists alike. Academic hubs like ARPANET mailing lists seeded Internet culture, while projects like GNU and Linux flourished in this open, collaborative environment. Today, the Internet stands in stark contrast to its early days. A few corporate tech giants operate data monopolies or quasi-monopolies, dictating the terms of participation to tightly controlled proprietary ecosystems.

This enclosure of the digital frontier represents not just a technical shift, but a political and economic one: a transition from a decentralized, participatory and inclusive network to largely an extractive one \cite{robinson2012nations}. However, Distributed Ledger Technologies (DLTs), a new technological evolution, could challenge the level playing field again.

\subsection{Distributed Ledger Technologies: new pathways for decentralization} 
\label{sec:intro_dlt}

Distributed Ledger Technologies are a relatively recent innovation integrating principles from game theory, cryptography, and distributed systems to solve the issue of \textit{double-spending}, that is, how to prevent that a valuable asset is spent multiple time when such asset is digital and the transaction involves two parties that do not trust each other~\cite{bohme2015bitcoin}. Blockchain technology is perhaps the most widely adopted type of DLT, with Bitcoin representing its first and most well-known application. Since its conception in 2008, it introduced a new economic paradigm by enabling digital payments to be settled in a decentralized way, without the need for an intermediary \cite{nakamoto2008bitcoin}.

In Bitcoin's design, users (or nodes) are connected in a peer-to-peer network and can transact native Bitcoin tokens with one another using pseudonymous addresses rather than accounts tied to verified identities.
Specialized users, known as miners, compete to earn the right to validate a new `block' of transactions by solving a computationally intensive cryptographic puzzle. The miner of a new block is rewarded by the platform in monetary units of cryptocurrency, but only in the case that all transactions in the block are valid. 
All users observe the new block and accept it only if all transactions are valid, which can be easily verified through cryptographic techniques; thus, consensus on the state of the system is reached without the oversight of an intermediary, aligning economic incentives with network security and integrity.

More recent advances in blockchain technology have led to more sophisticated platforms, like Ethereum, which not only facilitate peer-to-peer payments but also support decentralized applications (dApps) that allow users to trade, lend, borrow, and invest cryptocurrencies in a fully decentralized fashion \cite{werner2022sok}. This new financial ecosystem, known as Decentralized Finance (DeFi), relies on smart contracts -- software programs encapsulating functions and algorithms running on top of DLTs such as Ethereum -- that offer financial services without the need for an intermediary \cite{auer2024technology}. Notably, smart contracts also enable the issuance of non-native tokens, that is, cryptoassets\footnote{Henceforth, we will use interchangeably in this context the terms cryptocurrency, cryptoasset, digital asset, and token.} created on an existing blockchain and designed to address specific platform needs, such as representing ownership claims over underlying assets, serving particular functions like access credentials in a DeFi application, or even facilitating decision-making and the governance of decentralized Applications through collective voting on proposed changes to the dApp.

Blockchain's design inherently challenges centralization: anyone can participate in the network or validate transactions without approval from a central authority; it is accessible from anywhere with an internet connection and transcends national boundaries; transactions are censorship resistant and cannot be blocked or reversed by a central authority; no single entity controls the data record, rather, all users independently reach a consensus on the most recent state of the blockchain; thus, no single point of failure exists in the network, which remains active even if many nodes go offline; there isn't a central authority governing the Bitcoin protocol, and changes to it require broad consensus among participants.

Clearly, DLTs also face significant socio-economic and technical challenges. Mining, particularly in Bitcoin-like systems, demands substantial computational power and energy, resulting in high environmental costs and large carbon footprint~\cite{de2018bitcoin}. The lack of clear regulation facilitates market manipulation and illicit activity. Cryptoasset distribution is highly unequal and concentrated in the hands of a few. Technical limitations include scalability constraints and security vulnerabilities, especially when user participation is low. However, arguably, Bitcoin and other cryptocurrencies are among the most significant advancements in monetary and financial economics over the last two decades \cite{alvarez2023cryptocurrencies} and have the potential to transform existing payment and monetary systems \cite{bohme2015bitcoin}.
DLTs also have the potential to empower individuals by redistributing control from corporations or governments to communities, fundamentally altering how trust and authority are structured \cite{seidel2018questioning}. Decentralized Autonomous Organizations (DAOs) stand out as the main vehicle of this potential paradigm shift.

\section{Decentralized Autonomous Organizations (DAOs)}

Decentralized Autonomous Organizations (DAOs) have emerged in the last decade as a promising alternative to centralized structures. By leveraging blockchain technology, DAOs remove intermediaries and place decision-making in the hands of participants through automated smart contracts and token-based voting systems. This structure challenges the traditional top-down models of governance, offering a new form of organizational coordination that prioritizes transparency and collective input.

\begin{figure}[h!]
	\centering
	\includegraphics[width=0.95\linewidth]{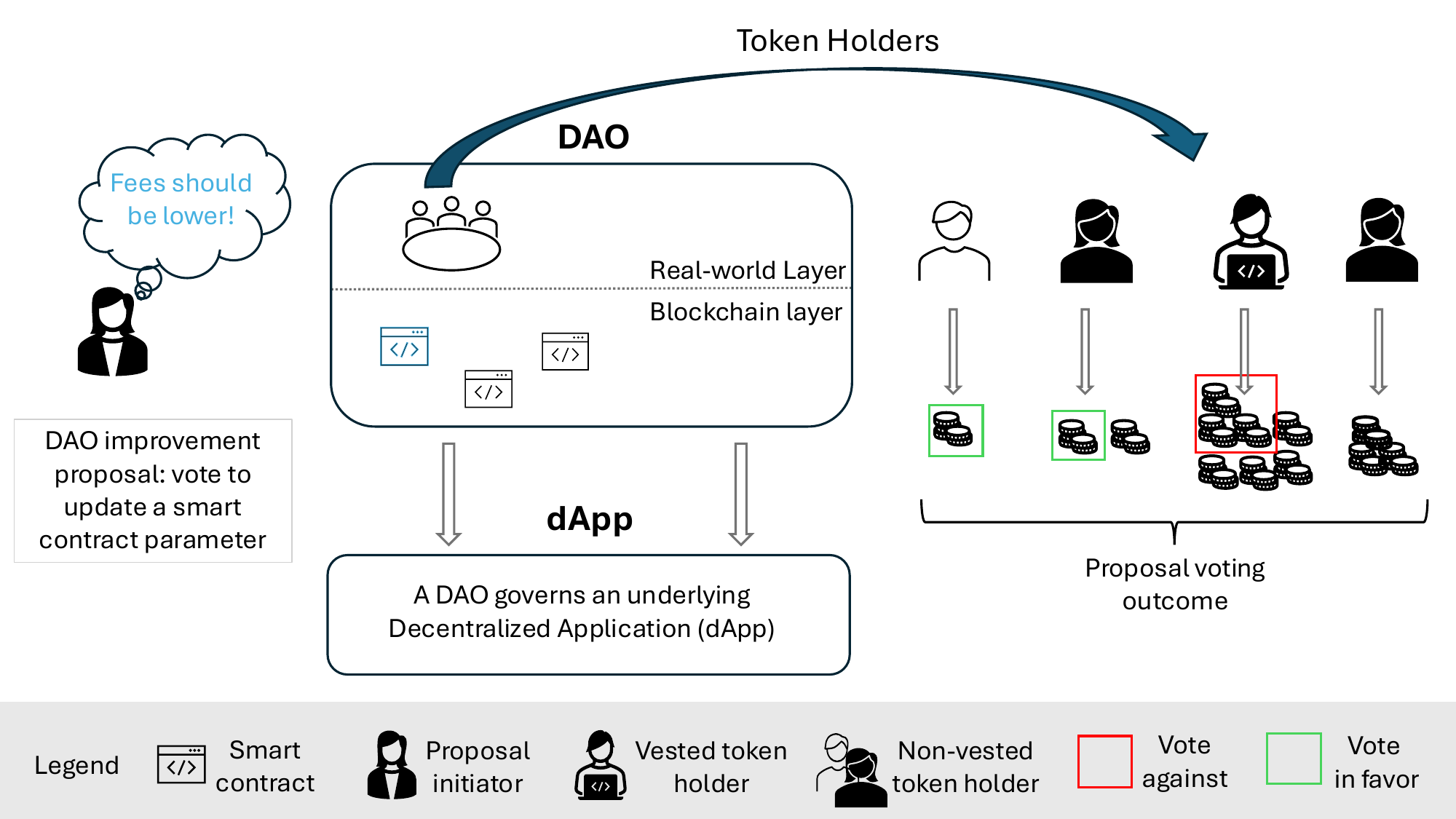}
	\caption{\textbf{Conceptual diagram of a DAO ecosystem}. The image illustrates the interaction between the key actors of a DAO: token holders (right), proposal initiators (left), and the DAO itself (center). The latter consists of a set of smart contracts implementing the logic on the blockchain, a governance structure---composed of all token holders---and potentially an underlying dApp that is managed by the DAO. The flow shows how suggested changes to the DAO structure or the underlying dApp, known as improvement proposals, are submitted and voted on with governance tokens. In this illustrative example, the proposer suggests modifying the parameter for the fees paid when utilizing the underlying dApp, which is controlled via the blue smart contract. Voters, who can either be vested (i.e., with a particular interest or role in the DAO) or non-vested users, can submit their preference in favor or against the proposal, or can decide not to exercise their voting right. If the proposal is accepted, the changes are implemented on the blockchain layer through changes on the smart contract. \textit{\small{Source: own elaboration.}}}
	\label{fig:DAO_concept}
\end{figure}

Figure~\ref{fig:DAO_concept} illustrates the main actors involved in DAOs and their interactions.
A DAO typically governs a decentralized application (dApp) and its associated smart contracts~\cite{zou2019smart}. The DAO itself is implemented on-chain as a suite of smart contracts embedding the operational logic intended by the developers.
This includes the purpose of the DAO and how it operates in practice.
Most DAOs use ``ERC-20'' tokens to represent voting power. ERC-20 tokens are a standard Ethereum smart contract that defines a set of widely accepted rules to represent and facilitate the transfer of value in an alternative form to that of the native cryptoasset, like Ether \cite{auer2024technology}. This includes equity tokens, representing a claim on an underlying asset, utility token, enabling access to specific functionalities (e.g. credentials), and governance tokens, which grant voting rights and decision-making power to users who become stakeholders in the governance structure of a DAO.

In DAOs, decision-making occurs through voting on so-called improvement proposals. These proposals can shape the development of the technical infrastructure~\cite{Uniswap2022deploy}, adjust parameters that influence economic incentives and system design~\cite{Compound2023migration,curve2020inflation}, or redirect the allocation of funds under the DAO’s control~\cite{Uniswap2023fees,Uniswap2023donation}. To take part in governance, users must hold governance tokens.
Voting occurs in two different forms, i.e. on-chain and off-chain. On-chain voting takes place directly on a DLT via smart contracts that encode the voting rules. In this setup, token holders may delegate their voting rights to another address, potentially controlled by a different party. While this method ensures high levels of security and transparency, it often suffers from high transaction fees, making it economically less viable~\cite{Dotan2023,Feichtinger2023}. In contrast, off-chain voting occurs on centralized platforms such as Snapshot~\cite{snapshot2023documentation,Wang2022b}, where only the final voting result is recorded on the blockchain. This approach is more cost-effective, user-friendly, and scalable but introduces greater centralization risks—such as the possibility of DAOs not honoring votes, parallel voting on multiple platforms, or reliance on databases that are not tamper-proof. 
In both cases, governance tokens are used to decide what is the outcome of the vote. DAO proposals define both a voting strategy, i.e. the
set of options published as part of a specific proposal, and the voting power, i.e. the weight associated with the amount of tokens cast by a user to vote. The most common voting scheme is one with a yes-no choice and follows the `one token, one vote' principle. 
Ultimately, the outcome of the proposal is decided by the entire community holding governance tokens.

At their core, DAOs aim to address fundamental issues with centralization: power imbalances, lack of transparency, and limited avenues for democratic participation. DAOs promise more equitable forms of governance by distributing power among stakeholders and embedding rules in the very code that drives organizational processes. This is in stark contrast to systems in which decisions are made by a small group of actors—be they corporate executives, political leaders, or institutional elites—whose actions may not align with the best interests of the broader community.
Beyond this, we note that the programmable nature of these governance rules revives longstanding debates about digital authority that emerged with the advent of cyberspace, but with a fundamental shift from Lessig's `code is law' principle~\cite{lessig2009code}, that indicates the process of incorporating legal rules into code, to the emerging `law is code' paradigm~\cite{de2018blockchain}, suggesting that code is becoming central not just in enforcing rules, but in creating them in the first place.

\subsection{Main Domains of Applicability of DAOs}

DAOs can be applied to any domain involving human interactions. Perhaps its most successful use case is for Decentralized Finance (DeFi) applications \cite{barbereau2023decentralised,auer2024technology}. In DeFi, DAOs manage all the parameters for decentralized financial protocols, such as lending and interest rates, risk parameters, and liquidity pools sizes. DeFI DAOs have been among the first to emerge into the spotlight in the summer of 2020, thanks to DeFi projects like Compound \cite{leshner2019compound}, SushiSwap \cite{sushiswap2022website}, and Uniswap \cite{adams2021uniswap} that released their governance tokens. These tokens bootstrapped communities and turned users into stakeholders. Protocols moved from being team-controlled to community-controlled via on-chain DAOs. Due to the explosive growth of DeFi on that period, on-chain activity surged leading to a spike in the price of various crypto assets involved in DeFi. Since then, prices have stabilized considerably, but DeFi Summer 2020 forever transformed DeFi users into DAO participants---redefining how protocols grow, govern, and distribute power.

DAOs go beyond financial applications. For instance, Decentralized Science (DeSci) is an emerging movement that seeks to transform how scientific research is funded, conducted, reviewed, and shared by leveraging blockchain-based tools and decentralized governance models \cite{weidener2025decentralization}. DeSci aims to address long-standing inefficiencies in traditional science—such as opaque peer review, limited funding access, and the enclosure of data and intellectual property—by promoting openness, collaboration, and community-driven resource allocation. Activities within the DeSci ecosystem are diverse: funding DAOs like VitaDAO allocate capital to underfunded or high-risk research; platforms like Molecule facilitate Intellectual Property (IP) marketplaces for licensing and collaboration; LabDAO builds infrastructure for sharing protocols and computational resources; and ResearchHub seeks to decentralize access to peer review and scientific discourse. The decentralization of power and resources in DeSci promises a more transparent, efficient, and inclusive knowledge production system.

DAOs are also often employed to manage creative commons. Collectives of creators use DAOs to coordinate creator rights, royalties, and community ownership of art, music, and intellectual property, often in the form of Non-Fungible Tokens (NFTs)\cite{balietti2025crypto}---a special digital asset that uniquely identifies ownable digital resources. These DAOs enable creators and their audiences to co-own and co-govern creative projects in a transparent, programmable way. For example, PleasrDAO is a collective that acquires culturally significant digital art and fractionalizes ownership among its members.  Similarly, Zora DAO is building infrastructure for creators to launch and sell media in a decentralized market. These DAOs reduce reliance on intermediaries and give artists more control over how their work is monetized and governed.

Directly related are DAOs governing guilds, game creators, or virtual economies, particularly in the metaverse and play-to-earn ecosystems. A prominent example is The Sandbox DAO, which oversees governance of the Sandbox metaverse, deciding over virtual land use, creator incentives, and in-game assets. Another is Yield Guild Games (YGG), a play-to-earn gaming guild DAO that invests in NFTs for blockchain games and lends them to players in exchange for revenue sharing. These DAOs coordinate large-scale participation in virtual economies, allowing players and developers to collaboratively shape the economic and governance structures of digital worlds.

Finally, there exist a number of DAOs focused on venture capital and asset management, where members pool capital to invest in startups, tokens, NFTs, and other blockchain-native assets. These investment DAOs offer a more transparent and participatory alternative to traditional venture capital. For instance, MetaCartel Ventures and The LAO are two early investment DAOs that provide funding to Web3 startups, with decisions made collectively by token holders. Another example is Orange DAO, a venture DAO formed by Y Combinator alumni to invest in early-stage crypto companies. Related to this are public goods DAOs such as Gitcoin DAO and Optimism Collective, which fund open-source infrastructure, decentralized tooling, and socially impactful projects using community-driven governance and novel mechanisms like quadratic funding or retroactive public goods funding. These DAOs blur the line between capital deployment and mission-driven community building.

\subsection{DAOs Features and Opportunities} 

DAOs represent an innovative institutional form that reconfigures traditional governance structures by embedding decision-making processes in open, programmable protocols executed by smart contracts. This shift enables a range of new opportunities to strengthen institutions across the dimensions of incentive alignment, coordination, resilience, inclusion, and accountability (see Table \ref{tab:dao_opportunities} for a summary with examples).

\begin{table}[h!]
\centering
\begin{tabular}{lp{0.2\textwidth}p{0.36\textwidth}p{0.30\textwidth}}
\hline
\textbf{\#} &  \textbf{Feature}  & \textbf{Description} & \textbf{Example DAOs} \\
\hline
(i) & Incentive alignment & Couples governance decisions with financial implications for token holders & \href{https://makerdao.com/}{MakerDAO (now Sky)}, \href{https://uniswap.com}{Uniswap} \\
\hline
(ii) & Global coordination & Enables borderless collaboration, fund raising, and collective action across domains & \href{https://gitcoin.co}{Gitcoin DAO}, \href{https://vita.io}{VitaDAO}, \href{https://metacartel.xyz}{MetaCartel Ventures}, \href{https://klimadao.finance}{KlimaDAO} \\
\hline
(iii) & Auditable Governance & Improves trust and accountability by recording all decisions and votes on-chain & \href{https://compound.finance/governance}{Compound DAO}, \href{https://aragon.org}{Aragon}, \href{https://ens.domains}{ENS DAO} \\
\hline
(iiii) & Broad Participation & Grants unrestricted access to investment, lending, and value creation  & \href{https://aave.com}{Aave}, \href{https://bitdao.io}{BitDAO}, \href{https://nexusmutual.io}{Nexus Mutual} \\
\hline
(v) & Evolutionary governance & Allows for custom rule-making, automation, and upgrades without centralized gatekeeping & \href{https://daohaus.club}{DAOhaus}, \href{https://aragon.org}{Aragon}, \href{https://compound.finance/governance}{Compound DAO} \\
\hline
(vi) &  Resilience in adversary context & Encourages decentralized operations and storage, reducing single points of failure & \href{https://ipfs.tech}{IPFS}, \href{https://filecoin.io}{Filecoin}, \href{https://ens.domains}{ENS DAO},  \\
\hline

\end{tabular}
\caption{\textbf{DAOs feaures across domains with examples.} }
\label{tab:dao_opportunities}
\end{table}

\subsubsection{Alignment: incentives for good governance}

DAOs embody a model of direct democracy in which governance rights are allocated in proportion to token ownership, creating a built-in mechanism for aligning incentives between stakeholders and decision-makers. By merging the roles of principals and agents, DAOs reduce the traditional agency problems that arise from the separation of ownership and control in corporate structures \cite{jensen1976theory, fama1983separation}. Token holders are not merely passive investors but active participants in governance whose financial interests are directly tied to the outcomes of their decisions. This alignment incentivizes them to act in ways that support the long-term success of the organization, promoting sustained engagement and collective accountability.

Incentive alignment is key for the governance of DeFi DAOs such as MakerDAO, Uniswap, or Compound. For instance, at MakerDAO, MKR holders vote on key risk parameters—like collateral types and debt ceilings—that directly affect the stability of the DAI stablecoin and the value of MKR itself; at Uniswap’s UNI holders decide whether to activate the protocol fee switch, a decision that requires balancing user adoption with potential revenue generation. The bottom line is that governance decisions have immediate financial implications for token holders, tightly coupling decision-making authority with economic incentives and long-term protocol performance.

\subsubsection{Coordination: collective actions and rapid capital allocation}

DAOs minimize coordination costs that typically hinder collective action. The automation of trust and execution through code can outperform centralized governance structures under specific conditions, enhancing the welfare of all stakeholders involved \cite{bena2023token}. This capacity for decentralized, large-scale coordination makes DAOs a powerful tool for organizing global communities around shared goals, specially when rapid capital allocation is required.

A textbook example of DAOs coordinating rapidly toward a common goal is ConstitutionDAO, which raised over \$40 million in days to bid on a rare copy of the U.S. Constitution for public display. Thousands of participants worldwide got mobilized in this direct democratic effort. Though the bid ultimately failed, the smart contract operated flawlessly, refunding contributors without delay and dissolving the DAO itself. This episode demonstrates the true potential of DAOs for targeted collective action.

\subsubsection{Accountability: transparent and auditable governance}

The operations of DAOs are recorded on public blockchains and governed by open-source smart contracts, enabling real-time auditing by both members and external observers. This radical transparency reduces information asymmetries and mitigates the risks associated with opaque decision-making. By making every transaction, proposal, and vote verifiable and publicly accessible, DAOs foster a governance environment built on trust, accountability, and verifiability \cite{wright2015blockchain, buterin2014daos}.

For example, Aragon DAO provides tools that enable organizations to run fully transparent governance processes, including real-time dashboards showing voting participation and treasury management. The Gitcoin DAO regularly publishes detailed reports of fund disbursement and community votes, reinforcing financial accountability to its contributors. These examples highlight how blockchain-based transparency combined with open governance mechanisms can enhance accountability in decentralized organizations.

\subsubsection{Inclusion: broad and permissionless participation}

DAOs enable the permissionless inclusion of a broad and diverse spectrum of stakeholders—ranging from users and developers to investors and contributors—in governance processes without traditional gatekeepers or centralized control. This open access lowers barriers to participation, allowing anyone with an interest and network access to engage directly in decision-making. For example, Gitcoin DAO empowers developers and community members worldwide to influence the allocation of funding for public goods. Similarly, MolochDAO fosters collaboration among Ethereum developers through open membership and collective decision-making. Such inclusivity fosters richer community engagement, encourages diverse perspectives, and helps build a more sustainable and resilient ecosystem \cite{li2021tokens, cardillo2023governance}. By embedding participation mechanisms into their governance structures, DAOs promote transparency, enhance trust, and strengthen organizational legitimacy.

\subsubsection{Responsiveness: evolution and programmability of governance}

Traditional organizations often suffer from structural inertia, making it difficult to adapt governance models or decision-making processes in response to changing environments \cite{kelly1991organizational}. In contrast, DAOs can evolve dynamically: they can upgrade or fork their governance logic, implement modular plugins (e.g., quadratic voting, conviction voting), and experiment with novel governance mechanisms at relatively low cost and with high speed \cite{reijers2020now, hassan2021modular}. Furthermore, DAOs can interoperate via on-chain protocols and shared standards, enabling DAO-to-DAO coordination, meta-governance, and composable substructures. For example, ecosystems such as Cosmos and Optimism have developed networks of interoperable DAOs that govern shared infrastructure while preserving organizational autonomy \cite{mushtaq2023composable, gill2022modular}.

\subsubsection{Resilience: support under adversarial contexts}

DAOs built on public blockchains are inherently more resistant to censorship, unilateral shutdown, or political interference. This resilience is particularly valuable for activist, journalistic, or dissident communities operating in adversarial contexts. By distributing governance and infrastructure across decentralized networks, DAOs can maintain operations even under external pressure \cite{zamfir2019censorship, hassan2021decentralized}. 

For instance, UkraineDAO and AssangeDAO leveraged this resilience to coordinate global fundraising and political messaging under hostile and fast-moving conditions. UkraineDAO raised millions in crypto donations during the war with Russia, while AssangeDAO supported WikiLeaks founder Julian Assange through NFT auctions. Similarly, IranDAO and HongKongDAO have explored decentralized models of resistance and mutual aid in regions with repressive governments. In such cases, DAOs function not only as financial tools but also as platforms for organizing and amplifying dissent, protected by their decentralized architecture.

\subsection{Challenges}

DAOs are not without challenges. Despite their decentralized ethos, many suffer from low participation rates, with governance decisions driven by a small subset of highly active or influential token holders. This can lead to de facto centralization, where “old-boy networks” or coordinated groups dominate outcomes. Additionally, technical complexities and voter fatigue may discourage wider engagement, limiting the democratizing potential of these organizations. Addressing these issues is crucial for realizing the full promise of decentralized governance.

\begin{table}[h!]
    \centering
    \begin{tabular}{lp{0.2\textwidth}p{0.33\textwidth}p{0.33\textwidth}}
        & Category & Mechanism & Negative consequences \\
        \hline
        (i) & Token concentration & Unequal token distribution, large holdings by insiders or exchanges & De facto centralization, dominance by wealthy actors, reduced fairness \\
        \hline
        (ii) & Participation & Low voter turnout, voter fatigue, technical barriers & Governance controlled by small active minority, erosion of democratic legitimacy \\
        \hline
        (iii) & Insider influence & Disproportionate power of contributors and vested users & Governance capture, insider trading, self-serving proposals \\
        \hline
        (iv) & Token trading & Vote buying/selling, strategic market manipulation & Market-driven governance distortions, exploitation of proposals for profit \\
        \hline
        (v) & Anonymity and coalitions & Pseudonymous actors, hidden alliances, conflicts of interest & Lack of accountability, self-dealing, vote manipulation, erosion of trust \\
        \hline
        (vi) & Technical and socio-technical limits & Overreliance on rule-based systems, suppression of ambiguity & Loss of diversity and personalization, risk of authoritarian dynamics \\
    \end{tabular}
    \caption{\textbf{Challenges of DAOs and decentralization}.}
    \label{tab:dao_challenges}
\end{table}

\subsubsection{Token concentration and centralization in DAOs}

Most DAOs use ERC-20 governance tokens to represent voting power. How tokens are initially distributed depends on each DAO's design. Whilst in most cases they are earned by users through participation, money supply and monetary policy are determined by the DAO creators. Perhaps not so surprisingly, this often leads to unbalances. Typically, a certain amount of tokens is initially allocated to the DAO treasury, contributors, and investors. Mechanisms such as airdrops, i.e. marketing strategies for expanding the user network based on the free distribution of tokens to current or potential users, further favor early participants and DAO members~\cite{makridis2023rise}. Moreover, the token distribution to users is typically proportional to the amount they invested, further rewarding wealthy users. 

Prior research has extensively documented that the on-chain ownership of governance tokens is, indeed, highly concentrated~\cite{Stroponiati2020governance,Jensen2021,Nadler2020,barbereau2022defi,Dotan2023}. Studies focusing specifically on DAO voting confirm that governance tokens and votes cast are highly concentrated \cite{Wang2022b,Laturnus2023}. 
Moreover, crypto trading platforms (CEXs) like Binance or Coinbase hold large amounts of governance tokens and, while they never exercised their voting rights, technically they are able to do so~\cite{Barbereau2023}.
A recent analysis of 342 DAOS by Cong et al.~\cite{cong2025centralized} shows that blockvoters, i.e. users with voting power exceeding 5\% of a proposal's total vote, control 76.2\% of the voting power in decisions.

\subsubsection{Low participation in voting}

A second major issue with DAOs is that users rarely exercise their voting rights. Studies show that the ratio between eligible and active voters has decreased over time, reaching 1 to 2\% in the latest analyzed voting outcomes~\cite{Barbereau2023}.
According to other studies, the participation rate in DAOs averaged around 6.3\%, a much lower rate than that observed in traditional corporations (70\%-80\%)~\cite{cong2025centralized}. Moreover, individuals who possess the potential power to alter outcomes rarely exercise it \cite{Feichtinger2023}. 
As an example, in major protocols like Uniswap and Compound, on average, less than 10\% of tokens are used to participate in voting processes~\cite{Fritsch2022a}.

\subsubsection{DAO vested users and the role of governance}

Voting power is concentrated in the hands of a few. Notably, researchers have also documented that DAO contributors or vested users -- including project owners, administrators, developers, and all other users with a relevant role in the creation of a DAO -- hold a disproportionate position in decision-making and are able to influence voting outcomes. 
Intuitively, contributors have higher incentives to be involved in the decision-making in DAOs. An analysis of the contributor involvement in DAO decision-making conducted on 872 DAOs~\cite{kitzler2024governance} shows that for 7.54\% DAOs, contributors have on average enough voting power to steer the governance alone, and they are significantly more likely to be found towards the center of the DAO governance ecosystem compared to non-vested users. 

Insiders might exploit their advantageous position to engage in opportunistic trading, with deep economic implications on the DAO ecosystem: Cong et al.~\cite{cong2025centralized} show that blockvoters are able to influence governance outcomes, and the managers of DAO improvement proposals conduct insider trading at a market-adjusted return of 9.5\% (while no significant short-term returns are obtained by other voters).

\subsubsection{DAO token trading} 

Governance tokens are cryptoassets and, therefore, carry a market value and can be traded both at centralized and decentralized exchanges. This creates a series of plausibly unintended consequences, including vote buying or selling and a disproportionate decision-making power of wealthy users: rather than following the more common approach in political voting of `one person, one vote', DAOs typically use the `one token, one vote' approach. Moreover, the tradability of governance tokens opens up avenues for strategic market manipulation---for example, a trader could short sell a token and then vote unfavorably in the DAO to drive the price lower, profiting from the induced decline.
Researchers have both identified evidence of abnormal trading activity before proposal creation and voting~\cite{cong2025centralized} and of majority shifts, i.e. trades large enough to swing the final outcome of the poll~\cite{kitzler2024governance}. 
A preliminary study reported examples of voters who held governance tokens for the duration of a single proposal life-cycle~\cite{Dotan2023}.

In mid-2024, Compound, a leading DeFi lending protocol, experienced a significant governance attack that exposed vulnerabilities inherent to trading of governance tokens. A prominent whale, known as ``Humpy,'' accumulated a substantial amount of COMP tokens from multiple wallets and leveraged this position to pass a proposal that directed 30\% of Compound’s reserves to a vault under their control. This move was widely perceived by the community as a governance attack, exploiting the protocol's decision-making mechanisms for personal gain. \footnote{\url{https://thedefiant.io/news/defi/compound-governance-attack-reveals-inherent-vulnerabilities-of-daos}}

\subsubsection{Anonymity versus accountability: coalitions and conflicts of interest}

One of the core features of DAOs is anonymity. Also this, however, presents significant challenges. 
When participants can propose, vote, and influence outcomes without revealing their identities, it becomes difficult to detect or prevent conflicts of interest. This lack of transparency can undermine trust, enable self-dealing, and make accountability nearly impossible, especially in cases where financial or strategic incentives are at play. Several high-profile incidents illustrate the risks: In April 2022, Beanstalk DAO was attacked by an anonymous user who stole \$182 millions from their funds\footnote{\url{https://medium.com/immunefi/hack-analysis-beanstalk-governance-attack-april-2022-f42788fc821e}}. In the Curve ecosystem, pseudonymous actors engaged in so-called \textit{gauge wars}, where they accumulated governance tokens to influence how token rewards were distributed. These actions were often coordinated off-chain and used to benefit projects they secretly controlled, raising concerns about vote-buying and hidden agendas\footnote{\url{https://www.gate.com/learn/articles/liquidity-wars-3-0-where-bribes-become-markets/9012}}.

Similar concerns about internal conflicts and opaque decision-making affected other DAOs governing SushiSwap, OlympusDAO, BadgerDAO DeFi protocols\footnote{\url{https://www.forbes.com/sites/tatianakoffman/2020/09/11/sushi-chef-returns-14m-apologizes-to-investors/}, \url{https://www.halborn.com/blog/post/explained-the-badgerdao-hack-december-2021}}.

Anonymous voters can create coalitions more easily. 
Kitzler et al.~\cite{kitzler2024governance} constructed a network capturing the user co-voting structure, and utilized the Louvain community detection method to identify communities of co-voting users. 
Two related works identify the existence of voters' coalitions in MakerDAO \cite{Sun2022,Sun2023}.
Finally,~\cite{cong2025centralized} show that conflicts of interest across different groups have a detrimental effect on the growth of a DAO.

\subsubsection{Technical and socio-technical limitations} 

Blockchain systems also face technical challenges, with scalability and performance issues being among the most prominent. These arise from the inherent computational inefficiency of consensus protocols, full ledger replication requirements, and energy-intensive proof-of-work mechanisms~\cite{Ruoti2019SoKBT}. On-chain DAO governance is subject to additional challenges due to its reliance on smart contracts, whose immutable code can contain vulnerabilities~\cite{Chaliasos2024}. Once deployed, these vulnerabilities become permanently embedded in the system, potentially causing severe consequences for the operation of the underlying application~\cite{Schaer2020} or organization~\cite{Feichtinger2024}.

More broadly, programmable or blockchain-based systems like DAOs are not designed to engage with qualities and values that cannot be well quantified, such as human dignity, creativity, affection, or passion. How these deeply human dimensions are preserved or transformed in such an ecosystem remains an open question. 
Similarly, a world where everyone agrees on a single, authoritative, and perfectly synchronized state of the world, like blockchains promise to do, risks producing a system that is not only limited but potentially dystopian, where diverse world views and ambiguities cannot exist. Not only is the multiplicity of perspectives at the basis of many human interactions, but its absence might pave the way to the emergence of authoritarian systems in the future.

Finally, it is worth noting that the rationale behind blockchain adoption builds on the belief that individuals are selfish, non-cooperative, and will exploit opportunities without ethical constraints. It distrusts human-run organizations, and instead places trust in a supposedly neutral, rule-based technology to manage interactions~\cite{parana2025inseparability}, despite the central role of social and institutional trust in enabling cooperation and stability within complex societies.

\section{DAOs and AI}

A distinctive feature of DAOs, compared to traditional organizations, is their programmable ability to evolve. This makes them apt to face current and future challenges, like those we highlighted in the previous section. 

A much-awaited development in the DAOs' ecosystem is the confluence with the recent trends in artificial intelligence (AI). AI technologies, with their capacity for analyzing large datasets, predicting outcomes, and automating processes, present the potential to augment significantly the functionality of DAOs. By integrating AI, DAOs could become even more adaptive and would gain the ability to automate complex processes that might otherwise be hindered by human limitations. For instance, the low participation rates discussed above would be eliminated by AI agents closely monitoring the DAOs for votes that are relevant for specific interests and informing users if there is a proposal deserving attention or even voting on behalf of a user according to pre-specified preferences (even on Sunday night or during summer holiday). Ailve DAO\footnote{\url{https://ethglobal.com/showcase/ailve-dao-9bi3b}.} is an attempt of proposing such an integration of AI and DAOs. It uses an AI agent to manage all proposals, including voting, automatically analyzing them through NLP and machine learning. Furthermore, their AI can control voting results and automatically disburse funds.

Yet, while the convergence of AI and DAOs holds transformative potential, it also raises profound concerns. AI systems, often opaque and difficult to understand, can introduce new risks—such as bias and accountability issues. In the eventuality that real human activity in DAOs is largely reduced in favor of AI agents, human actors would actually be left with poor control over agency. In fact, agents might not be doing their best interest for a number of reasons. First, changes in macro conditions might not be picked up by agents siloed in a restricted DAO ecosystem; second, human principals trusting the AI agent too much might likewise be oblivious to changes both in the macro environment and the DAO ecosystem. In either case, the principal fails to update their preferences or to communicate the update to the agent, potentially leading to undesirable outcomes. However, even when human principal and AI agent are perfectly aligned, AI agents could be gamed by malicious actors. In fact, it is known that AI systems may suffer from several security and privacy vulnerabilities \cite{yao2024survey} and fall prey to adversarial attacks such as jailbreaking \cite{liu2024jailbreaksurvey, qin2023masterkey,yu2024don} and prompt injection~\cite{liu2023prompt}.

The intersection of DAOs and AI must be approached with caution because---beyond individual mistakes---it also poses a systemic risk of a return to centralization. Potential risks of poorly integrated AI in DAOs include algorithmic bias skewing governance outcomes in favor of entrenched interests, automated decision-making that reduces human oversight and marginalizes dissenting voices, and the emergence of ``black box'' AI systems whose internal workings are inaccessible to participants. For example, if an AI system controlling voting weights or proposal prioritization is developed without community oversight, it could concentrate influence and undermine trust. In this context, it is important advocating for digital institutions that prioritize participatory design, transparent deliberation, and governance that emerges from diverse stakeholder collaboration. In this context, the development of public digital infrastructure---such as Polis and vTaiwan in Taiwan---can inform the evolution of DAOs from token-weighted voting systems toward more pluralistic, inclusive models of on-chain coordination \cite{weyl2024plurality}. 

For DAOs to extend the lessons of inclusive digital infrastructure like Polis and vTaiwan, there is a need for shared technical standards supporting pluralistic governance while integrating AI. The Agent-to-Agent (A2A) protocol provides a general framework for how AI agents can communicate and exchange tasks in a verifiable, inspectable way.\footnote{\url{https://a2a-protocol.org/specification}} By using A2A, DAOs can avoid ``black box'' systems and ensure that AI contributions remain under community oversight. Furthermore, for Ethereum-based DAOs---the vast majority of DAOs---this general framework is currently being refined by the ERC-8004 proposal on ``Trustless Agents.'' This ERC aims at establishing three on‑chain registries for Identity, Reputation, and Validation of actions of AI agents, enabling mechanisms for discovering and trusting agents in untrusted settings.\footnote{\url{https://eips.ethereum.org/EIPS/eip-8004}} Together, A2A and ERC-8004 show how technical protocols can complement inclusive design practices, helping DAOs integrate AI while preserving transparency, pluralism, and auditability.

\section{Conclusion}

Decision-making in both DAOs and AI systems shares a key feature: outcomes are often difficult to attribute to any single accountable individual or entity. As such, lessons from DAO governance and decision-making structures can provide valuable insights for developing AI frameworks that are more transparent and accountable. Looking ahead, DAOs may benefit from integrating emerging technologies such as artificial intelligence to enhance efficiency and coordination. At the same time, DAOs need to address persistent challenges within the crypto ecosystem---most notably, the excessive concentration of power in the hands of a few---to ensure that democratic practices are genuinely sustained.

In fact, the distribution of cryptocurrencies among users today is highly unequal. The Gini coefficient, a common measure of wealth inequality that ranges from 0 (perfect equality) to 1 (one individual owning everything), is very close to 1 for Bitcoin since its inception~\cite{kondor2014rich}. This has not changed over time, nor across cryptocurrencies~\cite{campajola2022evolution}. For comparison, the Gini index of the U.S. is currently around 0.42, and the metrics computed for cryptocurrencies do not account for the large population of individuals who do not own any cryptocurrencies at all. If power in DAOs is allocated through tokens with such unequal distributions, the decentralization premise breaks down, revealing what are essentially just `autonomous organizations' (AOs) in disguise.

However, the problem goes beyond unequal wealth distribution. Operations like updates to the protocol of a cryptocurrency or incident resolutions are managed by a restricted group of users. Previous research has documented that of all Ethereum Improvement Proposals (EIPs), i.e. the files formally suggesting changes or updates to the Ethereum blockchain, 68\% were proposed by ten individuals~\cite{fracassi2024decentralized}. A similar pattern has been documented for Bitcoin~\cite{gervais2014bitcoin} as well. These examples illustrate a core contradiction in the crypto ecosystem---including DAOs: while smart contracts may be self-executing and immutable, the social mechanisms behind them remain anything but autonomous.

If DAOs are neither decentralized nor autonomous, are they simply corporations on the blockchain? \footnote{This question is also raised by survey participants in ``DAOs: The New Coordination Frontier'': \url{https://www.bankless.com/the-ultimate-dao-report}} Even if originally envisioned as novel, democratized forms of organization and decision-making, many DAOs have been evolving in ways that mirror traditional institutions. In practice, they are often highly centralized---sometimes even more so than their conventional counterparts. Rather than eliminating intermediaries, DAOs might become a new type of virtual organization that introduces alternative forms of intermediation.

Still, DAOs represent a tremendous opportunity to confront the century-old problem of decision-making power concentrated in the hands of a few—arguably the most promising attempt in recent times. However, it is crucial that DAOs tackle this issue at its core, rather than merely disguising it in a new form. The DAOs of the future should not devolve into  technocratic or plutocratic forms of governance, nor resemble a timocracy, where ``rulers accumulate wealth in ways that are odds with the commons''~\cite{barbereau2023decentralised}. Otherwise, like the mythological Hydra---whose heads would grow again once cut---they risk becoming the very same monster they sought to slay.


\bibliographystyle{unsrt}

\bibliography{daos,centralization,sos,crowdsourcing,dlt,ai}

\begin{thebibliography}{10}

\bibitem{hoyt1941forces}
Homer Hoyt.
\newblock Forces of urban centralization and decentralization.
\newblock {\em American Journal of Sociology}, 46(6):843--852, 1941.

\bibitem{hopkins1965elite}
Keith Hopkins.
\newblock Elite mobility in the roman empire.
\newblock {\em Past and Present}, pages 12--26, 1965.

\bibitem{ko2013regional}
Chiu~Yu Ko and Tuan-Hwee Sng.
\newblock Regional dependence and political centralization in imperial china.
\newblock {\em Eurasian Geography and Economics}, 54(5-6):470--483, 2013.

\bibitem{wimmer2010rise}
Andreas Wimmer and Yuval Feinstein.
\newblock The rise of the nation-state across the world, 1816 to 2001.
\newblock {\em American Sociological Review}, 75(5):764--790, 2010.

\bibitem{kwon2024100}
Spencer~Y Kwon, Yueran Ma, and Kaspar Zimmermann.
\newblock 100 years of rising corporate concentration.
\newblock {\em American Economic Review}, 114(7):2111--2140, 2024.

\bibitem{trueblood2025misalignment}
Jennifer~S Trueblood, David~B Allison, Sarahanne~M Field, Ayelet Fishbach,
  Stefan~DM Gaillard, Gerd Gigerenzer, William~R Holmes, Stephan Lewandowsky,
  Dora Matzke, Mary~C Murphy, et~al.
\newblock The misalignment of incentives in academic publishing and
  implications for journal reform.
\newblock {\em Proceedings of the National Academy of Sciences},
  122(5):e2401231121, 2025.

\bibitem{merton_1968_matthew}
R.K. Merton.
\newblock {The Matthew Effect in Science. The Reward and Communication Systems
  of Science Are Considered.}
\newblock {\em Science}, 159(810):56--63, 1968.

\bibitem{xie_undemocracy_2014}
Y.~Xie.
\newblock "{Undemocracy}": Inequalities in science.
\newblock {\em Science}, 344(6186):809--810, 2014.

\bibitem{bettencourt2007growth}
Lu{\'\i}s~MA Bettencourt, Jos{\'e} Lobo, Dirk Helbing, Christian K{\"u}hnert,
  and Geoffrey~B West.
\newblock Growth, innovation, scaling, and the pace of life in cities.
\newblock {\em Proceedings of the national academy of sciences},
  104(17):7301--7306, 2007.

\bibitem{gabaix2009power}
Xavier Gabaix.
\newblock Power laws in economics and finance.
\newblock {\em Annu. Rev. Econ.}, 1(1):255--294, 2009.

\bibitem{sinha2006evidence}
Sitabhra Sinha.
\newblock Evidence for power-law tail of the wealth distribution in india.
\newblock {\em Physica A: Statistical Mechanics and its Applications},
  359:555--562, 2006.

\bibitem{adamic2000power}
Lada~A Adamic and Bernardo~A Huberman.
\newblock Power-law distribution of the world wide web.
\newblock {\em science}, 287(5461):2115--2115, 2000.

\bibitem{diamond1999guns}
Jared~M Diamond and Doug Ordunio.
\newblock {\em Guns, germs, and steel}, volume 521.
\newblock Books on Tape New York, 1999.

\bibitem{rosenthal2001managing}
Uriel Rosenthal, Arjen Boin, and Louise~K Comfort.
\newblock {\em Managing crises: Threats, dilemmas, opportunities}.
\newblock Charles C Thomas Publisher, 2001.

\bibitem{robinson2012nations}
James~A Robinson and Daron Acemoglu.
\newblock {\em Why nations fail: The origins of power, prosperity and poverty}.
\newblock Profile London, 2012.

\bibitem{hong_diversity_vs_experts_2004}
L.~Hong and S.E. Page.
\newblock Groups of diverse problem solvers can outperform groups of
  high-ability problem solvers.
\newblock {\em Proceedings of the National Academy of Sciences (PNAS)},
  101(46):16385--16389, 2004.

\bibitem{kavadias_diversity_brainstorming_2009}
S.~Kavadias and S.C. Sommer.
\newblock The effects of problem structure and team diversity on brainstorming
  effectiveness.
\newblock {\em Management Science}, 55(12):1899--1913, 2009.

\bibitem{moch1977size}
Michael~K Moch and Edward~V Morse.
\newblock Size, centralization and organizational adoption of innovations.
\newblock {\em American sociological review}, pages 716--725, 1977.

\bibitem{brown1993centralized}
Duncan Brown.
\newblock Centralized control or decentralized diversity: A guide for matching
  compensation with company strategy and structure.
\newblock {\em Compensation \& Benefits Review}, 25(5):47--52, 1993.

\bibitem{bayram2023hub}
Vedat Bayram, Bar{\i}{\c{s}} Y{\i}ld{\i}z, and M~Saleh Farham.
\newblock Hub network design problem with capacity, congestion, and stochastic
  demand considerations.
\newblock {\em Transportation Science}, 57(5):1276--1295, 2023.

\bibitem{true2019punctuated}
James~L True, Bryan~D Jones, and Frank~R Baumgartner.
\newblock Punctuated-equilibrium theory: explaining stability and change in
  public policymaking.
\newblock In {\em Theories of the Policy Process, Second Edition}, pages
  155--187. Routledge, 2019.

\bibitem{lanier2014owns}
Jaron Lanier.
\newblock {\em Who owns the future?}
\newblock Simon and Schuster, 2014.

\bibitem{berners1999weaving}
Tim Berners-Lee.
\newblock {\em Weaving the Web: The original design and ultimate destiny of the
  World Wide Web by its inventor}.
\newblock Harper San Francisco, 1999.

\bibitem{bohme2015bitcoin}
Rainer B{\"o}hme, Nicolas Christin, Benjamin Edelman, and Tyler Moore.
\newblock Bitcoin: Economics, technology, and governance.
\newblock {\em Journal of economic Perspectives}, 29(2):213--238, 2015.

\bibitem{nakamoto2008bitcoin}
Satoshi Nakamoto.
\newblock Bitcoin: A peer-to-peer electronic cash system.
\newblock 2008.

\bibitem{werner2022sok}
Sam Werner, Daniel Perez, Lewis Gudgeon, Ariah Klages-Mundt, Dominik Harz, and
  William Knottenbelt.
\newblock Sok: Decentralized finance (defi).
\newblock In {\em Proceedings of the 4th ACM Conference on Advances in
  Financial Technologies}, pages 30--46, 2022.

\bibitem{auer2024technology}
Raphael Auer, Bernhard Haslhofer, Stefan Kitzler, Pietro Saggese, and Friedhelm
  Victor.
\newblock The technology of decentralized finance (defi).
\newblock {\em Digital Finance}, 6(1):55--95, 2024.

\bibitem{de2018bitcoin}
Alex De~Vries.
\newblock Bitcoin's growing energy problem.
\newblock {\em Joule}, 2(5):801--805, 2018.

\bibitem{alvarez2023cryptocurrencies}
Fernando Alvarez, David Argente, and Diana Van~Patten.
\newblock Are cryptocurrencies currencies? bitcoin as legal tender in el
  salvador.
\newblock {\em Science}, 382(6677):eadd2844, 2023.

\bibitem{seidel2018questioning}
Marc-David~L Seidel.
\newblock Questioning centralized organizations in a time of distributed trust.
\newblock {\em Journal of Management Inquiry}, 27(1):40--44, 2018.

\bibitem{zou2019smart}
Weiqin Zou, David Lo, Pavneet~Singh Kochhar, Xuan-Bach~Dinh Le, Xin Xia, Yang
  Feng, Zhenyu Chen, and Baowen Xu.
\newblock Smart contract development: Challenges and opportunities.
\newblock {\em IEEE Transactions on Software Engineering}, 47(10):2084--2106,
  2019.

\bibitem{Uniswap2022deploy}
Uniswap.
\newblock Improvement proposal: deploy uniswap v3 on bnb chain, 2022.
\newblock Available at:
  \url{https://gov.uniswap.org/t/rfc-update-deploy-uniswap-v3-1-0-3-0-05-0-01-on-bnb-chain-binance/19734}.

\bibitem{Compound2023migration}
Compound.
\newblock Improvement proposal: Compound v2 to v3 migration phase 1, 2023.
\newblock Available at:
  \url{https://compound.finance/governance/proposals/152}.

\bibitem{curve2020inflation}
Curve.
\newblock Improvement proposal: reduction to crv token inflation, 2020.
\newblock Available at:
  \url{https://gov.curve.fi/t/discussion-reduction-to-crv-inflation/851}.

\bibitem{Uniswap2023fees}
Uniswap.
\newblock Improvement proposal: Making protocol fees operational, 202.
\newblock Available at:
  \url{https://gov.uniswap.org/t/making-protocol-fees-operational/21198}.

\bibitem{Uniswap2023donation}
Uniswap.
\newblock Improvement proposal: Uniswap dao can help 80k uni to turkiye to
  relief after the big earthquake disaster, 2023.
\newblock Available at:
  \url{https://gov.uniswap.org/t/governance-proposal-uniswap-dao-can-help-80k-uni-to-turkiye-to-relief-after-the-big-earthquake-disaster/20758}.

\bibitem{Dotan2023}
Maya Dotan, Aviv Yaish, Hsin-Chu Yin, Eytan Tsytkin, and Aviv Zohar.
\newblock The {Vulnerable} {Nature} of {Decentralized} {Governance} in {DeFi}.
\newblock Technical report, August 2023.
\newblock arXiv:2308.04267 [cs] type: article.

\bibitem{Feichtinger2023}
Rainer Feichtinger, Robin Fritsch, Yann Vonlanthen, and Roger Wattenhofer.
\newblock The {Hidden} {Shortcomings} of ({D}){AOs} -- {An} {Empirical} {Study}
  of {On}-{Chain} {Governance}.
\newblock Technical report, February 2023.
\newblock arXiv:2302.12125 [cs] type: article.

\bibitem{snapshot2023documentation}
Snapshot.
\newblock Documentation.
\newblock Technical report, 2023.
\newblock Available at \url{https://docs.snapshot.org/}.

\bibitem{Wang2022b}
Qin Wang, Guangsheng Yu, Yilin Sai, Caijun Sun, Lam~Duc Nguyen, Sherry Xu, and
  Shiping Chen.
\newblock An {Empirical} {Study} on {Snapshot} {DAOs}.
\newblock Technical report, November 2022.
\newblock arXiv:2211.15993 [cs] type: article.

\bibitem{lessig2009code}
Lawrence Lessig.
\newblock {\em Code: And other laws of cyberspace}.
\newblock ReadHowYouWant. com, 2009.

\bibitem{de2018blockchain}
Primavera De~Filippi and Samer Hassan.
\newblock Blockchain technology as a regulatory technology: From code is law to
  law is code.
\newblock {\em arXiv preprint arXiv:1801.02507}, 2018.

\bibitem{barbereau2023decentralised}
Tom Barbereau, Reilly Smethurst, Orestis Papageorgiou, Johannes Sedlmeir, and
  Gilbert Fridgen.
\newblock Decentralised finance’s timocratic governance: The distribution and
  exercise of tokenised voting rights.
\newblock {\em Technology in Society}, 73:102251, 2023.

\bibitem{leshner2019compound}
Robert Leshner and Geoffrey Hayes.
\newblock Compound: The money market protocol.
\newblock Technical report, 2019.
\newblock Available at \url{https://bit.ly/3ioWOjW}.

\bibitem{sushiswap2022website}
Sushiswap.
\newblock Documentation.
\newblock Tech. rep., 2022.
\newblock Available at \url{https://docs.sushi.com/}.

\bibitem{adams2021uniswap}
Hayden Adams, Noah Zinsmeister, Moody Salem, River Keefer, and Dan Robinson.
\newblock Uniswap v3 core.
\newblock Technical report, 2021.
\newblock Available at \url{https://uniswap.org/whitepaper-v3.pdf}.

\bibitem{weidener2025decentralization}
Lukas Weidener and Bence Luk{\'a}cs.
\newblock The decentralization of science and the emergence of decentralized
  science.
\newblock {\em Available at SSRN 5094597}, 2025.

\bibitem{balietti2025crypto}
Stefano Balietti, Can Celebi, and David Tercero-Lucas.
\newblock From crypto to nfts: Identifying the new wave of digital investors.
\newblock {\em International Review of Financial Analysis}, 104:104172, 2025.

\bibitem{jensen1976theory}
Michael~C Jensen and William~H Meckling.
\newblock Theory of the firm: Managerial behavior, agency costs and ownership
  structure.
\newblock {\em Journal of Financial Economics}, 3(4):305--360, 1976.

\bibitem{fama1983separation}
Eugene~F Fama and Michael~C Jensen.
\newblock Separation of ownership and control.
\newblock {\em Journal of Law and Economics}, 26(2):301--325, 1983.

\bibitem{bena2023token}
Jan Bena and Shiqi Zhang.
\newblock Token-based decentralized governance, data economy and platform
  business model.
\newblock {\em Available at SSRN 4248492}, 2023.

\bibitem{wright2015blockchain}
Aaron Wright and Primavera De~Filippi.
\newblock Blockchain technology: Beyond bitcoin.
\newblock {\em Communications of the ACM}, 58(9):38--39, 2015.

\bibitem{buterin2014daos}
Vitalik Buterin.
\newblock Daos, dacs, das and more: An incomplete terminology guide.
\newblock {\em Ethereum Blog}, 2014.
\newblock Available at:
  \url{https://blog.ethereum.org/2014/05/06/daos-dacs-das-and-more}.

\bibitem{li2021tokens}
Linda Li, William Mann, and Sundar Sankaraguruswamy.
\newblock Tokens and governance.
\newblock {\em NBER Working Paper No. 29131}, 2021.

\bibitem{cardillo2023governance}
Giovanni Cardillo, Alessandro Rossi, and Vlad Zamfir.
\newblock Governance in daos: Decentralization, participation, and performance.
\newblock {\em Journal of Institutional and Theoretical Economics}, 2023.
\newblock Forthcoming.

\bibitem{kelly1991organizational}
Dawn Kelly and Terry~L Amburgey.
\newblock Organizational inertia and momentum: A dynamic model of strategic
  change.
\newblock {\em Academy of management journal}, 34(3):591--612, 1991.

\bibitem{reijers2020now}
Wessel Reijers and Mark Coeckelbergh.
\newblock Now the code runs itself: On-chain governance and the future of
  decentralized coordination.
\newblock {\em Philosophy \& Technology}, 33(3):409--424, 2020.

\bibitem{hassan2021modular}
Samer Hassan.
\newblock Modular governance in daos: Patterns and best practices, 2021.
\newblock DAO Research Collective White Paper.

\bibitem{mushtaq2023composable}
Sarah Mushtaq.
\newblock Composable governance: An architecture for interoperable daos, 2023.
\newblock Available at:
  \url{https://medium.com/@mushtaq/composable-governance}.

\bibitem{gill2022modular}
James Gill.
\newblock Modular daos and composable organizations, 2022.
\newblock Available at: \url{https://banklessdao.substack.com/p/modular-daos}.

\bibitem{zamfir2019censorship}
Vlad Zamfir.
\newblock Censorship resistance in blockchain governance, 2019.
\newblock Available at:
  \url{https://vitalik.ca/general/2019/05/09/governance.html}.

\bibitem{hassan2021decentralized}
Samer Hassan, Primavera De~Filippi, et~al.
\newblock Decentralized governance: An exploration of daos, 2021.
\newblock Available at: \url{https://arxiv.org/abs/2102.03476}.

\bibitem{makridis2023rise}
Christos~A Makridis, Michael Fr{\"o}wis, Kiran Sridhar, and Rainer B{\"o}hme.
\newblock The rise of decentralized cryptocurrency exchanges: Evaluating the
  role of airdrops and governance tokens.
\newblock {\em Journal of Corporate Finance}, 79:102358, 2023.

\bibitem{Stroponiati2020governance}
Katerina Stroponiati, Ilya Abugov, Yiannis Varelas, Kostas Stroponiatis,
  Modesta Jurgeleviciene, and Yashoda Savanth.
\newblock Decentralized governance in defi: Examples and pitfalls.
\newblock Technical report, DappRadar, 2020.

\bibitem{Jensen2021}
Johannes~Rude Jensen, Victor von Wachter, and Omri Ross.
\newblock How {Decentralized} is the {Governance} of {Blockchain}-based
  {Finance}: {Empirical} {Evidence} from four {Governance} {Token}
  {Distributions}.
\newblock Technical report, February 2021.
\newblock arXiv:2102.10096 [q-fin] type: article.

\bibitem{Nadler2020}
Matthias Nadler and Fabian Schär.
\newblock Decentralized {Finance}, {Centralized} {Ownership}? {An} {Iterative}
  {Mapping} {Process} to {Measure} {Protocol} {Token} {Distribution}.
\newblock {\em arXiv:2012.09306 [cs, econ, q-fin]}, December 2020.
\newblock arXiv: 2012.09306.

\bibitem{barbereau2022defi}
Tom~Josua Barbereau, Reilly Smethurst, Orestis Papageorgiou, Alexander Rieger,
  and Gilbert Fridgen.
\newblock Defi, not so decentralized: The measured distribution of voting
  rights.
\newblock In {\em Proceedings of the Hawaii International Conference on System
  Sciences 2022}, page~10, 2022.

\bibitem{Laturnus2023}
Valerie Laturnus.
\newblock The {Economics} of {Decentralized} {Autonomous} {Organizations}.
\newblock Technical Report 4320196, Rochester, NY, January 2023.

\bibitem{Barbereau2023}
Tom Barbereau, Reilly Smethurst, Orestis Papageorgiou, Johannes Sedlmeir, and
  Gilbert Fridgen.
\newblock Decentralised finance's timocratic governance: {The} distribution and
  exercise of tokenised voting rights.
\newblock {\em Technology in Society}, page 102251, April 2023.

\bibitem{cong2025centralized}
Lin~William Cong, Daniel Rabetti, Charles~CY Wang, and Yu~Yan.
\newblock Centralized governance in decentralized organizations.
\newblock {\em Available at SSRN 5168660}, 2025.

\bibitem{Fritsch2022a}
Robin Fritsch, Marino Müller, and Roger Wattenhofer.
\newblock Analyzing {Voting} {Power} in {Decentralized} {Governance}: {Who}
  controls {DAOs}?
\newblock Technical report, April 2022.
\newblock arXiv:2204.01176 [cs] type: article.

\bibitem{kitzler2024governance}
Stefan Kitzler, Stefano Balietti, Pietro Saggese, Bernhard Haslhofer, and
  Markus Strohmaier.
\newblock The governance of decentralized autonomous organizations: A study of
  contributors’ influence, networks, and shifts in voting power.
\newblock In {\em International Conference on Financial Cryptography and Data
  Security}, pages 313--330. Springer, 2024.

\bibitem{Sun2022}
Xiaotong Sun, Charalampos Stasinakis, and Georigios Sermpinis.
\newblock Decentralization illusion in {DeFi}: {Evidence} from {MakerDAO}.
\newblock Technical report, March 2022.
\newblock arXiv:2203.16612 [cs, q-fin] type: article.

\bibitem{Sun2023}
Xiaotong Sun, Xi~Chen, Charalampos Stasinakis, and Georgios Sermpinis.
\newblock Voter {Coalitions} and democracy in {Decentralized} {Finance}:
  {Evidence} from {MakerDAO}.
\newblock Technical report, June 2023.
\newblock arXiv:2210.11203 [cs, q-fin] version: 4 type: article.

\bibitem{Ruoti2019SoKBT}
Scott Ruoti, Ben Kaiser, Arkady Yerukhimovich, Jeremy Clark, and Robert~K.
  Cunningham.
\newblock Sok: Blockchain technology and its potential use cases.
\newblock {\em ArXiv}, abs/1909.12454, 2019.

\bibitem{Chaliasos2024}
Stefanos Chaliasos, Marcos~Antonios Charalambous, Liyi Zhou, Rafaila
  Galanopoulou, Arthur Gervais, Dimitris Mitropoulos, and Ben Livshits.
\newblock Smart {Contract} and {DeFi} {Security} {Tools}: {Do} {They} {Meet}
  the {Needs} of {Practitioners}?
\newblock In {\em Proceedings of the 46th {IEEE}/{ACM} {International}
  {Conference} on {Software} {Engineering}}, pages 1--13, February 2024.
\newblock arXiv:2304.02981 [cs].

\bibitem{Schaer2020}
Fabian Schär.
\newblock Decentralized {Finance}: {On} {Blockchain}- and {Smart}
  {Contract}-based {Financial} {Markets}.
\newblock Technical Report 3571335, Rochester, NY, March 2020.

\bibitem{Feichtinger2024}
Rainer Feichtinger, Robin Fritsch, Lioba Heimbach, Yann Vonlanthen, and Roger
  Wattenhofer.
\newblock {SoK}: {Attacks} on {DAOs}.
\newblock In Rainer Böhme and Lucianna Kiffer, editors, {\em 6th {Conference}
  on {Advances} in {Financial} {Technologies} ({AFT} 2024)}, volume 316 of {\em
  Leibniz {International} {Proceedings} in {Informatics} ({LIPIcs})}, pages
  28:1--28:27, Dagstuhl, Germany, 2024. Schloss Dagstuhl – Leibniz-Zentrum
  für Informatik.

\bibitem{parana2025inseparability}
Edemilson Paran{\'a}.
\newblock The inseparability between technological domination and financial
  hegemony in contemporary capitalism.
\newblock {\em Retheorising capitalism}, page 263, 2025.

\bibitem{yao2024survey}
Yifan Yao, Jinhao Duan, Kaidi Xu, Yuanfang Cai, Zhibo Sun, and Yue Zhang.
\newblock A survey on large language model (llm) security and privacy: The
  good, the bad, and the ugly.
\newblock {\em High-Confidence Computing}, page 100211, 2024.

\bibitem{liu2024jailbreaksurvey}
Xiaosen Liu, Zhengyan Zhang, Yiming Cao, et~al.
\newblock Jailbreak attacks and defenses against large language models: A
  survey.
\newblock {\em arXiv preprint arXiv:2407.04295}, 2024.

\bibitem{qin2023masterkey}
Chen Qin, Zifan Liu, Qian Zou, et~al.
\newblock Masterkey: Automated jailbreak across multiple large language model
  chatbots.
\newblock {\em arXiv preprint arXiv:2307.08715}, 2023.

\bibitem{yu2024don}
Zhiyuan Yu, Xiaogeng Liu, Shunning Liang, Zach Cameron, Chaowei Xiao, and Ning
  Zhang.
\newblock Don't listen to me: understanding and exploring jailbreak prompts of
  large language models.
\newblock In {\em 33rd USENIX Security Symposium (USENIX Security 24)}, pages
  4675--4692, 2024.

\bibitem{liu2023prompt}
Yi~Liu, Gelei Deng, Yuekang Li, Kailong Wang, Zihao Wang, Xiaofeng Wang,
  Tianwei Zhang, Yepang Liu, Haoyu Wang, Yan Zheng, et~al.
\newblock Prompt injection attack against llm-integrated applications.
\newblock {\em arXiv preprint arXiv:2306.05499}, 2023.

\bibitem{weyl2024plurality}
E.~Glen Weyl, Audrey Tang, and the Plurality Research~Network.
\newblock {\em Plurality: Technology and the Future of Democracy}.
\newblock Plurality Press, 2024.
\newblock Open source and collaboratively written.

\bibitem{kondor2014rich}
D{\'a}niel Kondor, M{\'a}rton P{\'o}sfai, Istv{\'a}n Csabai, and G{\'a}bor
  Vattay.
\newblock Do the rich get richer? an empirical analysis of the bitcoin
  transaction network.
\newblock {\em PloS one}, 9(2):e86197, 2014.

\bibitem{campajola2022evolution}
Carlo Campajola, Raffaele Cristodaro, Francesco~Maria De~Collibus, Tao Yan,
  Nicolo' Vallarano, and Claudio~J Tessone.
\newblock The evolution of centralisation on cryptocurrency platforms.
\newblock {\em arXiv preprint arXiv:2206.05081}, 2022.

\bibitem{fracassi2024decentralized}
Cesare Fracassi, Moazzam Khoja, and Fabian Sch{\"a}r.
\newblock Decentralized crypto governance? transparency and concentration in
  ethereum decision-making.
\newblock {\em Transparency and Concentration in Ethereum Decision-Making
  (January 10, 2024)}, 2024.

\bibitem{gervais2014bitcoin}
Arthur Gervais, Ghassan~O Karame, Vedran Capkun, and Srdjan Capkun.
\newblock Is bitcoin a decentralized currency?
\newblock {\em IEEE security \& privacy}, 12(3):54--60, 2014.

\end{thebibliography}


\end{document}